\documentclass[prd,a4paper,twocolumn,showpacs,nofootinbib]{revtex4}
\usepackage[a4paper, hdivide={1.91cm,,1.165cm}, vdivide={1.83cm,,4.5cm}]{geometry}

\usepackage{amstext,amssymb,amsmath}
\usepackage[english]{babel}
\usepackage[ansinew]{inputenc}
\usepackage{graphicx}
\usepackage{units}
\usepackage{hyperref}

\allowdisplaybreaks

\hypersetup{
    pdftitle={Leptogenesis with Lepton-Number-Violating Dirac Neutrinos},
    pdfauthor={Julian Heeck}
}

\begin{document}

\let \Lold \L
\def \L {\mathcal{L}} 
\let \epsilonold \epsilon
\def \epsilon {\varepsilon} 

\newcommand{\tr}{\mathrm{tr}}
\newcommand{\hc}{\mathrm{h.c.}}
\newcommand{\im}{\mathrm{Im}}

\title{Leptogenesis with Lepton-Number-Violating Dirac Neutrinos}

\author{Julian Heeck}
\email[Electronic address: ]{julian.heeck@mpi-hd.mpg.de}
\affiliation{Max-Planck-Institut f\"ur Kernphysik, Saupfercheckweg 1, 69117 Heidelberg, Germany}


\begin{abstract}

Dirac neutrinos with lepton-number-violating interactions can give rise to a new leptogenesis mechanism. In its simplest renormalizable realization, based on a gauged $B-L$ symmetry spontaneously broken by four units, the decay of a new scalar creates an asymmetry in the right-handed neutrinos. A neutrinophilic two-Higgs-doublet model converts this asymmetry to the baryons, provides a natural explanation of the small neutrino masses, and can lead to an effective number of relativistic degrees of freedom of $N_\mathrm{eff} \simeq 3.29$ due to the entropy-suppressed contribution of the right-handed neutrinos.

\end{abstract}

\maketitle


\section{Introduction}

Introducing heavy right-handed Majorana neutrinos to the Standard Model (SM) of particle physics can provide a solution to several phenomena beyond the SM. For one, the active neutrinos acquire naturally small Majorana masses through the seesaw mechanism~\cite{seesaw}, in accordance with neutrino oscillation experiments. Furthermore, the apparent asymmetry of matter over antimatter is explained via leptogenesis~\cite{leptogenesis} through the $CP$-violating leptonic decay of the heavy sterile neutrinos in the early Universe. The resulting lepton asymmetry is then partially transferred to the baryon sector via sphalerons, i.e.~nonperturbative processes violating baryon plus lepton number ($B+L$)~\cite{sphalerons}. See Ref.~\cite{Fong:2013wr} for a recent review.

One of the inherent predictions of this framework is the Majorana nature of the light neutrinos---the neutrino is its own antiparticle and $B-L$ is a broken symmetry. This allows, most importantly, for neutrinoless double beta decay~\cite{Rodejohann:2011mu}, which has yet to be observed. Until then, the question of the neutrino nature is still open, and neutrinos might just be Dirac particles like all other known fermions. 
In order to generate the necessary sub-eV masses for the neutrinos, the Yukawa couplings $y$ to the SM Higgs boson $H$ then have to be tiny, $y\sim m_\nu/\langle H\rangle \lesssim 10^{-11}$. Over the years, many models have been brought forward to explain these small couplings in a more natural way~\cite{Roncadelli:1983ty,Murayama:2002je}, and even a leptogenesis mechanism with Dirac neutrinos was proposed~\cite{dirac_leptogenesis} (see also~\cite{Murayama:2002je}). 
This so-called neutrinogenesis makes use of the fact that the Yukawa couplings are too small to thermalize the right-handed neutrinos (RHNs) in the early Universe. The $CP$-violating, but $(B-L)$-conserving, decay of some heavy particle can then create an asymmetry for left-handed leptons $\Delta_L$ that is canceled by an asymmetry for the right-handed leptons $\Delta_R = -\Delta_L$. With all particles in equilibrium, these asymmetries would be washed out by the sphalerons; however, due to the small couplings, any $\Delta_R$ stored in the RHNs will be smuggled past the sphalerons, which consequently create a nonzero baryon asymmetry only from $\Delta_L$.

An interesting and very different route to motivate light Dirac neutrinos has been discussed in Refs.~\cite{Wang:2006jy,Gabriel:2006ns,Davidson:2009ha}, where a second Higgs doublet $H_2$ is introduced, which couples exclusively to neutrinos~\cite{Ma:2000cc}. A small vacuum expectation value (VEV), say $\langle H_2 \rangle \sim \unit[1]{eV}$, is then the reason for small neutrino masses, while the Yukawa couplings can be large. This leads to distinctive collider signatures~\cite{Davidson:2010sf}, but also makes standard neutrinogenesis impossible.
In this paper we will provide a new kind of Dirac leptogenesis, which relies on thermalized RHNs and therefore works for the neutrinophilic two-Higgs-doublet solution of small Dirac masses.
Our mechanism uses the recently introduced framework of lepton-number-violating (LNV) Dirac neutrinos~\cite{Heeck:2013rpa} to create a lepton asymmetry from the $CP$-violating decay of a heavy particle.\footnote{Prior to Ref.~\cite{Heeck:2013rpa}, it was already mentioned in Ref.~\cite{Chen:2012jg} that LNV Dirac neutrinos could lead to interesting effects in the early Universe.} As such, the mechanism is actually more reminiscent of standard leptogenesis than neutrinogenesis, even though it contains Dirac neutrinos.

\section{LNV Dirac Neutrinos}

Let us briefly review the simplest model for lepton-number-violating Dirac neutrinos, brought forward in Ref.~\cite{Heeck:2013rpa}. We work with a gauged $B-L$ symmetry, three RHNs $\nu_R \sim -1$, one scalar $\phi\sim 4$ to break $B-L$, and one scalar $\chi \sim -2$ as a mediator, all of which are singlets under the SM gauge group. The Lagrangian takes the form
\begin{align}
\begin{split}
 \L \ &= \L_\mathrm{SM} +\L_\mathrm{kinetic} +\L_{Z'}- V(H,\phi,\chi) \\
&\quad + \left( y_{\alpha\beta} \overline{L}_\alpha H \nu_{R,\beta} + \frac{1}{2}\kappa_{\alpha\beta} \chi \, \overline{\nu}_{R,\alpha} \nu_{R,\beta}^c  +\hc\right) ,
\end{split}
\label{eq:model}
\end{align}
$H$ being the SM Higgs doublet.
If $\chi$ does not acquire a VEV, the neutrinos will be Dirac particles with mass matrix $M_{\alpha\beta} = y_{\alpha\beta} |\langle H \rangle| = U \mathrm{diag} (m_1^\nu, m_2^\nu, m_3^\nu) V_R^\dagger$. The smallness of neutrino masses is, in this simple model, a result of very small couplings, $|y_{\alpha\beta}| \lesssim 10^{-11}$. The symmetric Yukawa coupling matrix $\kappa_{\alpha \beta} = \kappa_{\beta \alpha}$ is nondiagonal and complex in general, which is important for our leptogenesis application in the next section.
The scalar potential takes the form
\begin{align}
\begin{split}
 &V(H,\phi,\chi) \equiv \sum_{X= H, \phi,\chi} \left(\mu_X^2 |X|^2 + \lambda_X |X|^4\right) \\
 &\quad + \sum_{\substack{X, Y= H, \phi,\chi\\ X\neq Y}}  \frac{\lambda_{X Y}}{2} |X|^2 |Y|^2
 - \mu\left( \phi \chi^2 + \hc \right) ,
\end{split}
\label{eq:potential}
\end{align}
with symmetric couplings $\lambda_{X Y} = \lambda_{Y X}$.
Choosing the structure $\mu_H^2,\mu_\phi^2 < 0 < \mu_\chi^2$, one can easily realize a potential with minimum $\langle \chi\rangle =0$, $\langle H\rangle\neq 0 \neq \langle \phi \rangle$, which breaks $SU(2)_L\times U(1)_Y \times U(1)_{B-L}$ to $U(1)_\mathrm{EM}\times \mathbb{Z}_4^L$. An exact $\mathbb{Z}_4^L$ symmetry remains, under which leptons transform as $\ell \to -i \,\ell$ and $\chi \to - \chi$, making the neutrinos Dirac particles but still allowing for $\Delta L = 4$ LNV processes.\footnote{Conservation of lepton number modulo $n > 2$ as a means to forbid Majorana neutrino masses was also mentioned in Ref~\cite{Witten:2000dt}.}
The crucial $\mu$ term in the potential will induce a mass splitting between the real scalars $\Xi_j$ in $\chi = (\Xi_1 + i\, \Xi_2)/\sqrt{2}$:
\begin{align}
 m_{1}^2 = m_c^2 - 2\mu\langle \phi\rangle\,, &&
 m_{2}^2 = m_c^2 + 2\mu\langle \phi\rangle\,,
\end{align}
where $m_c$ is a common mass term that is of no importance here. 
Since the $\Xi_j$ can decay in either $\nu_R \nu_R$ or $\nu_R^c \nu_R^c$, lepton number is clearly violated, even though our model has Dirac neutrinos. The scalars also induce a $\Delta L =4$ scattering $\nu_R\nu_R \to \nu_R^c\nu_R^c$ and potentially mediate neutrinoless quadruple beta decay $(A,Z)\to (A,Z+4) + 4\, e^-$~\cite{Heeck:2013rpa}.

Let us note that the $\mathbb{Z}_4^L$ symmetry left over after breaking $B-L$ could also be used as the stabilizing symmetry behind dark matter. For example, an even $B-L$ charge for a newly introduced Dirac fermion would make it exactly stable, because all other fermions in the SM$+\nu_R$ carry odd $B-L$ charge.

\section{Dirac Leptogenesis}

As seen above, neutrinos are Dirac particles in our model, yet $B-L$ is broken, which makes possible a real Dirac lepto\emph{genesis}, where a lepton asymmetry is created by the $CP$-violating $\Delta (B-L) = 4$ decay of some heavy particle. In order for this to work, the decay has to take place after $B-L$ breaking and before the electroweak phase transition (EWPT), so that sphalerons can convert the lepton asymmetry to the baryons (assuming $\Delta B = 0$ as induced in our model).

For a simple realization, we use the framework of the previous section and add second copies of both $\chi$ and $H$, both $\chi_j$ without VEVs.
Below the $B-L$ breaking scale, $\chi_1$ and $\chi_2$ now split into four real scalars $\Xi_j$, with decay channels $\nu_{R,\alpha} \nu_{R,\beta}$ and $\nu_{R,\alpha}^c \nu_{R,\beta}^c$. $\chi_2$ is necessary to obtain $CP$ violation in these decays (depicted in Fig.~\ref{fig:leptogenesis}), as we will see below. The out-of-equilibrium decay of the lightest $\Xi_j$ then has all the necessary qualitative features to create an asymmetry $\Delta_{\nu_R}$ in the right-handed neutrinos. This in itself would not suffice for baryogenesis, as the sphalerons do not see the right-handed $\Delta_{\nu_R}$, and the Higgs Yukawa couplings $y \sim m_\nu/\langle H_1\rangle$ from Eq.~\eqref{eq:model} are too small to efficiently convert $\Delta_{\nu_R}$ to the left-handed lepton doublets. This is where the second Higgs doublet $H_2$ comes in, as it can have large enough Yukawa couplings $w_{\alpha\beta} \overline{L}_\alpha H_2 \nu_{R,\beta}$ to thermalize $\nu_R$ and transfer $\Delta_{\nu_R} \rightarrow \Delta_{L}$. From 
there, sphalerons take over to convert $\Delta_L$ to the baryons $\Delta_B$ in the usual leptogenesis fashion (see e.g.~Ref.~\cite{Fong:2013wr} for a review).

The second Higgs doublet $H_2$ will be chosen to be neutrinophilic, i.e.~with a small VEV~\cite{Davidson:2009ha}. While this is not strictly necessary for our version of Dirac leptogenesis---for example, a VEV-less $H_2$ with large Yukawas would work as well, the neutrinos gaining mass via $H_1$---it is the most interesting two-Higgs-doublet model~\cite{Branco:2011iw} for our purposes, as it additionally sheds light on the small neutrino masses.
To this effect, let us mention briefly how the neutrinophilic nature of $H_2$ can be realized in our context. Following Ref.~\cite{Gabriel:2006ns}, we impose an additional global $\mathbb{Z}_2$ symmetry (or a $U(1)$ as in Ref.~\cite{Davidson:2009ha}) under which only $H_2$ and $\nu_R$ are charged, forbidding all $H_2$ Yukawa couplings except $w_{\alpha\beta} \overline{L}_\alpha H_2 \nu_{R,\beta}$. The new symmetry is broken softly by a term $\mu_{12}^2 H_1^\dagger H_2$ in the scalar potential. A small $\mu_{12}^2$ is technically natural and will induce a small VEV for $H_2$, $\langle H_2 \rangle/\langle H_1 \rangle = \mu_{12}^2/M_{H_2}^2$, which gives naturally small Dirac neutrino masses $M_{\alpha\beta} = w_{\alpha\beta} |\langle H_2 \rangle|$. We stress that our additional $B-L$ symmetry and scalars, compared to Refs.~\cite{Gabriel:2006ns,Davidson:2009ha}, in no way complicate or interfere with this realization of a neutrinophilic $H_2$, so we will not go into any more details.

\begin{figure}[tb]
\setlength{\abovecaptionskip}{-1ex}
	\begin{center}
		\includegraphics[width=0.43\textwidth]{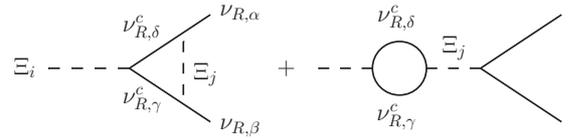}
	\end{center}
		\caption{$CP$-violating vertex and self-energy loop corrections to the LNV decay $\Xi_i \rightarrow \nu_{R, \alpha} \nu_{R, \beta}$ relevant for leptogenesis.}
	\label{fig:leptogenesis}
\end{figure}

After these qualitative statements, let us delve into a more quantitative analysis of our leptogenesis mechanism. The scalar potential for $\phi$, $H_{1,2}$ and $\chi_{1,2}$ is more involved than before (Eq.~\eqref{eq:potential}), but the only qualitatively new terms are 
\begin{align}
 V(\phi, H_{1,2},\chi_{1,2})\  \supset \ m_{12}^2\, \overline{\chi}_1 \chi_2 + \mu_{12}\, \phi\, \chi_1 \chi_2 + \hc,
\label{eq:scalar_mixing}
\end{align}
as they lead to a mixing of the four real fields $\Xi_j$ contained in $\chi_{1,2}$ after breaking $B-L$. The $4\times 4$ mass matrix for the $\Xi_j$ is not particularly illuminating, and a diagonalization just redefines the couplings $\kappa^j_{\alpha\beta}$ to the RHNs (see Eq.~\eqref{eq:model}). Since the resulting couplings are the only relevant ones for leptogenesis, we can skip all these steps and just work with four real scalar fields $\Xi_j$ with masses $m_{j}$ and complex symmetric Yukawa couplings $V^j_{\alpha\beta} = V^j_{\beta\alpha}$,
\begin{align}
 \L \  \supset \  \frac{1}{2} V^j_{\alpha \beta}\, \Xi_j \overline{\nu}_{R,\alpha} \nu^c_{R,\beta} +  \frac{1}{2} \overline{V}^j_{\alpha \beta}\, \Xi_j \overline{\nu}_{R,\alpha}^c \nu_{R,\beta} \,,
\label{eq:effective_yukawas}
\end{align}
where implicit sums are understood and $\overline{V}^j_{\alpha\beta} \equiv (V^j_{\alpha\beta})^*$.

The $Z'$ interactions will keep the SM particles and the new scalars and RHNs in equilibrium above $T_{Z'} \simeq (\sqrt{g_*} \langle \phi \rangle^4/M_\mathrm{Pl})^{1/3}$, $g_* \simeq 100$ being the effective number of degrees of freedom at temperature $T$ and $M_\mathrm{Pl}\simeq \unit[10^{19}]{GeV}$ the Planck mass. Below $T_{Z'}$, the real scalars $\Xi_j$ will only be coupled to the SM via the Higgs portal (assumed to be small for simplicity) and the RHN interactions from Eq.~\eqref{eq:effective_yukawas}. The out-of-equilibrium condition for the decay of the lightest $\Xi_i$ then reads
\begin{align}
 \Gamma (\Xi_i \rightarrow \nu_R \nu_R, \nu_R^c \nu_R^c) \ll H(T \sim m_{i}) \simeq 1.66\, \sqrt{g_*} \frac{m_{i}^2}{M_\mathrm{Pl}} \,,
\end{align}
$H(T)$ being the Hubble expansion rate of the Universe at temperature $T$ (not to be confused with the Higgs fields $H_j$).
As with the bulk of leptogenesis models, this condition is most naturally fulfilled for very heavy decaying particles, as can be seen by inserting the total decay rate $\Gamma (\Xi_i) = \tr (\overline{V}^i V^i)\, m_{i}/4\pi$, leading to
\begin{align}
 \tr (\overline{V}^i V^i)/10^{-6} \ll m_{i}/\unit[10^{11}]{GeV} \,,
\label{eq:out_of_equilibrium}
\end{align}
which can be satisfied with either small Yukawa couplings or large masses, in complete analogy to the standard leptogenesis with heavy right-handed Majorana neutrinos.

Assuming the out-of-equilibrium condition~\eqref{eq:out_of_equilibrium} to be satisfied, the decay of the lightest $\Xi_i$ then leads to a $CP$ asymmetry due to interference of tree-level and one-loop diagrams (Fig.~\ref{fig:leptogenesis}):
\begin{align}
 \epsilon_i \equiv 2 \, \frac{\Gamma \left(\Xi_i \rightarrow \nu_{R} \nu_{R} \right) - \Gamma \left(\Xi_i \rightarrow \nu_{R}^c \nu_{R}^c \right)}{\Gamma \left(\Xi_i \rightarrow \nu_R \nu_R \right) + \Gamma \left(\Xi_i \rightarrow \nu_R^c \nu_R^c \right)} \,,
\end{align}
where we already summed over flavor indices and included a factor of $2$ because two RHNs are created per decay.
A straightforward calculation yields the asymmetries from the vertex ($\epsilon^\mathrm{v}$) and self-energy correction ($\epsilon^\mathrm{s}$): 
\begin{align}
\begin{split}
 \epsilon^\mathrm{v}_i &= \frac{1}{4\pi} \frac{1}{\tr (\overline{V}^i V^i)} \sum_{k\neq i} F (\eta_k)\, \im \left[ \tr \left( \overline{V}^i V^k \overline{V}^i V^k \right) \right] ,\\
 \epsilon^\mathrm{s}_i &= -\frac{1}{24\pi} \frac{1}{\tr (\overline{V}^i V^i)} \sum_{k\neq i} G (\eta_k)\, \im \left[ \left\{ \tr \left( \overline{V}^i V^k \right) \right\}^2 \right] ,
\end{split}
\label{eq:asymmetries}
\end{align}
with $\eta_k \equiv m_{i}^2/m_{k}^2 < 1$ and the functions
\begin{align}
\begin{split}
 F (x) &\equiv \frac{x- \log (1+x)}{x} =  \frac{x}{2} + \mathcal{O} \left( x^2 \right) ,\\
 G (x) &\equiv \frac{x}{1-x} =  x + \mathcal{O} \left( x^2 \right) .
\end{split}
\end{align}
As quick cross-checks, one can easily verify that the $k=i$ contribution to the sums in Eq.~\eqref{eq:asymmetries} vanishes because the trace of a hermitian matrix is real. 
One can also convince oneself that the second $\chi_2$ is indeed necessary for the $CP$ asymmetry, as the couplings of just one field $\chi = (\Xi_1 + i \Xi_2)/\sqrt{2}$ would lead to the Yukawa-coupling relation $V^2 =i V^1$ and ultimately $\epsilon^\mathrm{s}=0=\epsilon^\mathrm{v}$.
Let us consider one last limiting case before we move on: Neglecting the $\chi_1$--$\chi_2$ mixing terms in the scalar potential~\eqref{eq:scalar_mixing} gives $\chi_1 = (\Xi_1 + i \Xi_2)/\sqrt{2}$, $\chi_2 =(\Xi_3+i \Xi_4)/\sqrt{2}$ and the relations $V^2 = i V^1$ and $V^4 = i V^3$. Assuming $\Xi_1$ to be the lightest of the four scalars, $\Xi_2$ does not contribute to $\epsilon$ by the argument given above. The contributions of $\Xi_3$ and $\Xi_4$ are opposite in sign, so that $\epsilon^\mathrm{v}\propto F(\eta_3) - F(\eta_4)$ and $\epsilon^\mathrm{s}\propto G(\eta_3) - G(\eta_4)$. The asymmetry therefore vanishes for $m_{3} = m_{4}$, as it should, because this would imply $B-L$ conservation.

Compared to other leptogenesis scenarios, the asymmetries from vertex and self-energy corrections in our model depend on different flavor parameters---even in the unflavored case---because $\tr (A^2) \neq (\tr A)^2$ for a general matrix $A$. The asymmetries are nevertheless qualitatively reminiscent of standard leptogenesis, with the same rough behavior $\epsilon \sim 10^{-7} (\eta/10^{-2}) (V/10^{-2})^2$---ignoring the complex matrix structure of $V$ and assuming a hierarchy $\eta_k \ll 1$. A low-scale resonant leptogenesis is of course also possible in our framework, but goes beyond the scope of this paper.

The total lepton asymmetry, i.e.~the RHN number density $n_{\nu_R}$ relative to the entropy density $s = (2\pi^2/45) g_* T^3$, is then given by
\begin{align}
 Y_{\nu_R} \equiv \frac{n_{\nu_R}}{s} \sim \frac{\epsilon^\mathrm{v}_i+\epsilon^\mathrm{s}_i }{g_*} \, .
\end{align}
Since we assume equilibrium of the SM particles with the RHNs as well as the sphalerons, we can use chemical potentials to describe the plasma. (Note that $B-L$ is effectively conserved once the $\Xi_j$ have dropped out.) Consequently, the chemical potential for the RHNs has to be added to the usual set of equations~\cite{Harvey:1990qw}, resulting in the equilibrium condition $3 B + L = 0$, or
\begin{align}
 Y_B = \frac{1}{4}\, Y_{B-L}\,, && Y_L = -\frac{3}{4}\, Y_{B-L}\,,
\end{align}
for three generations (and an arbitrary number of Higgs doublets), compared to $Y_B = \frac{28}{79} \, Y_{B-L}$ for standard leptogenesis with one Higgs doublet. The condition $3 B + L = 0$ can also be understood with the help of Ref.~\cite{Kartavtsev:2005rs}, where it was pointed out that $3 B +L$ vanishes if only left-handed fermions and the sphalerons are in equilibrium. Since we introduce fully thermalized right-handed partners to all left-handed fermions, it is no surprise that $3 B + L = 0$ remains valid.

With all of the above, it should be clear that our LNV Dirac neutrinos can accommodate the observed baryon asymmetry $Y_B\sim 10^{-10}$ in this novel leptogenesis scenario. We refrain from a parameter scan, as the Yukawa couplings $V^j$ and masses $m_{j}$ are in any way hardly constrained by other processes or related to other observables, at least for the very heavy $\Xi_j$ considered here. This leptogenesis mechanism is testable nonetheless, because it requires additional interactions for the RHNs.
Let us therefore discuss the last crucial piece of the puzzle: the thermalization of the RHNs.

The $\nu_R$ asymmetry needs to be transferred to the left-handed sector before the EWPT in order to generate the baryon asymmetry of the Universe. Correspondingly, we need stronger-than-usual interactions for the RHNs, in our case by means of the second Higgs doublet $H_2$ in $w_{\alpha\beta} \overline{L}_\alpha H_2 \nu_{R,\beta}$. At temperatures above the electroweak scale, the interaction rates go with $w^2 T$, which equilibrates the RHNs if $w \gtrsim 10^{-8}$~\cite{dirac_leptogenesis}. This does not lead to problems because, below the EWPT, the interaction rate drastically changes its form; the charged Higgs $H_2^+$, for example, mediates an $\ell^+ \ell^- \leftrightarrow \overline{\nu}_R \nu_R$ scattering with rate $w^4 T^5/m_{H_2^+}^4$, i.e.~suppressed by the mass. The RHN decoupling temperature $T^\mathrm{dec}_{\nu_R}$ is then given by the condition
\begin{align}
 w^4 \left( T^\mathrm{dec}_{\nu_R} \right)^5/m_{H_2^+}^4 \sim H \left( T^\mathrm{dec}_{\nu_R} \right) ,
\label{eq:nuRdecoupling}
\end{align}
at least for large $w$.
If the RHNs decouple before the left-handed neutrinos, i.e.~$T^\mathrm{dec}_{\nu_R} > T^\mathrm{dec}_{\nu_L} \sim \unit[1]{MeV}$, the RHN contribution to the effective number of relativistic degrees of freedom $N_\mathrm{eff}$ will be diluted~\cite{Davidson:2009ha}:
\begin{align}
 N_\mathrm{eff} \simeq 3 + 3 \left[ g_* (T^\mathrm{dec}_{\nu_L})/g_* (T^\mathrm{dec}_{\nu_R}) \right]^{4/3}  .
\end{align}
We have $g_* \left(T^\mathrm{dec}_{\nu_L}\right) = 43/4$, and recent Planck data constrain $N_\mathrm{eff} = 3.30 \pm 0.27$ at $68\%$ C.L.~\cite{Ade:2013zuv} (dependent on the combination of data sets). The RHNs therefore have to decouple before the QCD phase transition, $T^\mathrm{dec}_{\nu_R} > 150$--$\unit[300]{MeV}$, which yields, with Eq.~\eqref{eq:nuRdecoupling}, a bound on the Yukawa couplings~\cite{Davidson:2009ha}:
\begin{align}
 |w| \lesssim \frac{1}{30} \left( \frac{m_{H_2^+}}{\unit[100]{GeV}}\right) \left(\frac{1/\sqrt{2}}{|U_{\ell i}|}\right) .
\end{align}
Earlier decoupling is of course possible, but we always expect some contribution of the RHNs to $N_\mathrm{eff}$, namely, $3.14\lesssim N_\mathrm{eff} \lesssim 3.29$ for $\unit[150]{MeV} \lesssim T^\mathrm{dec}_{\nu_R} \lesssim \unit[200]{GeV}$, assuming only SM degrees of freedom.
These values can even explain the long-standing deviation of the best-fit value of $N_\mathrm{eff}$ from the SM value $3.046$, as recently emphasized in Ref.~\cite{Anchordoqui:2012qu}.
Consequently, the second Higgs doublet $H_2$ puts the RHNs in equilibrium above the EWPT to generate the baryon asymmetry, and then it naturally decouples them to satisfy and ameliorate cosmological constraints. Taking the flavor structure of the Yukawa couplings $w_{\alpha\beta}$ into account will modify the discussion a bit, but that goes beyond the scope of this paper. We refer to Refs.~\cite{Davidson:2009ha,Davidson:2010sf} for a detailed discussion of the phenomenology of the neutrinophilic $H_2$, which is still valid for our extension with lepton-number-violating Dirac neutrinos.

\section{Conclusion}
\label{sec:conclusion}

Dirac neutrinos with lepton-number-violating interactions make possible a new way to create a lepton asymmetry in the early Universe. In the simplest model presented here, this asymmetry resides in the right-handed neutrino sector and requires a second Higgs doublet to transfer it to the left-handed leptons and ultimately baryons. 
If the second doublet couples exclusively to neutrinos, its small vacuum expectation value can in addition provide a natural explanation for the smallness of the neutrino masses without invoking small Yukawa couplings. 
The unavoidable partial thermalization of the right-handed neutrinos contributes to the relativistic degrees of freedom in perfect agreement with the persisting observational hints. Together with the ensuing collider phenomenology of the second Higgs doublet and, of course, the predicted absence of neutrinoless double beta decay, this model can be falsified in current and upcoming experiments.

\begin{acknowledgments}
The author gratefully thanks Tibor Frossard and Werner Rodejohann for discussions and comments on the manuscript.
This work was supported by the Max Planck Society in the project MANITOP and by the IMPRS-PTFS.
\end{acknowledgments}


\end{document}